\documentclass[twocolumn,preprintnumbers,aps,pra,superscriptaddress,amsmath,amssymb]{revtex4-1}

\usepackage{graphicx}% Include figure files
\usepackage{dcolumn}% Align table columns on decimal point
\usepackage{bm}
\usepackage{bbm}
\usepackage{color}
\usepackage{multirow}
\usepackage{physics}

\begin{document}

\title{Spin properties of dense near-surface ensembles of nitrogen-vacancy centres in diamond}

\author{J.-P. Tetienne} 
\email{jtetienne@unimelb.edu.au}
\affiliation{School of Physics, The University of Melbourne, VIC 3010, Australia}	
	
\author{R. W. de Gille}
\affiliation{School of Physics, The University of Melbourne, VIC 3010, Australia}
	
\author{D. A. Broadway}
\affiliation{School of Physics, The University of Melbourne, VIC 3010, Australia}	
\affiliation{Centre for Quantum Computation and Communication Technology, School of Physics, The University of Melbourne, VIC 3010, Australia}

\author{T. Teraji}
\affiliation{National Institute for Materials Science, Tsukuba, Ibaraki 305-0044, Japan}

\author{S. E. Lillie}
\affiliation{School of Physics, The University of Melbourne, VIC 3010, Australia}	
\affiliation{Centre for Quantum Computation and Communication Technology, School of Physics, The University of Melbourne, VIC 3010, Australia}

\author{J. M. McCoey}
\affiliation{School of Physics, The University of Melbourne, VIC 3010, Australia}

\author{N. Dontschuk}
\affiliation{School of Physics, The University of Melbourne, VIC 3010, Australia}	
\affiliation{Centre for Quantum Computation and Communication Technology, School of Physics, The University of Melbourne, VIC 3010, Australia}

\author{L. T. Hall}
\affiliation{School of Physics, The University of Melbourne, VIC 3010, Australia}

\author{A. Stacey}
\affiliation{School of Physics, The University of Melbourne, VIC 3010, Australia}	
\affiliation{Centre for Quantum Computation and Communication Technology, School of Physics, The University of Melbourne, VIC 3010, Australia}

\author{D. A. Simpson}
\email{simd@unimelb.edu.au}
\affiliation{School of Physics, The University of Melbourne, VIC 3010, Australia}

\author{L. C. L. Hollenberg}
\affiliation{School of Physics, The University of Melbourne, VIC 3010, Australia}
\affiliation{Centre for Quantum Computation and Communication Technology, School of Physics, The University of Melbourne, VIC 3010, Australia}

\date{\today}
	
\begin{abstract}
	
We present a study of the spin properties of dense layers of near-surface nitrogen-vacancy (NV) centres in diamond created by nitrogen ion implantation. The optically detected magnetic resonance contrast and linewidth, spin coherence time, and spin relaxation time, are measured as a function of implantation energy, dose, annealing temperature and surface treatment. To track the presence of damage and surface-related spin defects, we perform in situ electron spin resonance spectroscopy through both double electron-electron resonance and cross-relaxation spectroscopy on the NV centres. We find that, for the energy ($4-30$~keV) and dose ($5\times10^{11}-10^{13}$~ions/cm$^2$) ranges considered, the NV spin properties are mainly governed by the dose via residual implantation-induced paramagnetic defects, but that the resulting magnetic sensitivity is essentially independent of both dose and energy. We then show that the magnetic sensitivity is significantly improved by high-temperature annealing at $\geq1100^\circ$C. Moreover, the spin properties are not significantly affected by oxygen annealing, apart from the spin relaxation time, which is dramatically decreased. Finally, the average NV depth is determined by nuclear magnetic resonance measurements, giving $\approx10$-17~nm at 4-6 keV implantation energy. This study sheds light on the optimal conditions to create dense layers of near-surface NV centres for high-sensitivity sensing and imaging applications.

\end{abstract}

\maketitle

\section{Introduction}

Over the last decade, the nitrogen-vacancy (NV) centre in diamond has become a leading solid-state quantum system for magnetic sensing and imaging applications \cite{Doherty2013,Rondin2014,Schirhagl2014}. While single NV centres located near the diamond surface provide higher spatial resolution, down to a few nanometres  \cite{Maletinsky2011,Mamin2013,Staudacher2013,Tetienne2014,Lovchinsky2016,Wood2017}, near-surface layers of high-density NV centres allow faster imaging through parallel acquisition, with a spatial resolution limited by the diffraction of light ($\approx300$~nm) \cite{Steinert2013,LeSage2013,Simpson2016,Tetienne2017,Simpson2017}. Thin layers of NV centres in diamond can be created mainly by two methods: (i) Chemical vapour deposition (CVD) growth with nitrogen incorporated into the gas phase \cite{Ohno2012,Ohashi2013,Kleinsasser2016}, and (ii) post-growth nitrogen ion implantation \cite{Rabeau2006,Steinert2010,Pham2011}. The CVD method generally results in NV centres with longer spin coherence times -- hence enhanced magnetic sensitivity -- as compared to implanted NV centres, due to reduced lattice damage, but suffers from a limited NV density \cite{Kleinsasser2016}. For applications such as nuclear magnetic resonance (NMR) imaging \cite{DeVience2015} and electron paramagnetic resonance (EPR) imaging \cite{Steinert2013,Simpson2017}, which require near-surface ($\lesssim20$~nm) layers of NV centres, nitrogen implantation thus remains the preferred method, despite inferior per-NV sensitivities. Here, we study the spin properties of dense near-surface layers of implanted NV centres as a function of implantation energy, dose, annealing temperature and surface treatment, with the aim of identifying the optimal conditions for NMR/EPR imaging applications. 

Magnetometry with the NV centre in diamond generally relies on optically detected magnetic resonance (ODMR) of the NV centre's electron spin. For photon shot noise limited measurements, the sensitivity to dc magnetic fields \cite{Taylor2008,Degen2008,Rondin2014} is given by
\begin{eqnarray} \label{Eq:eta_dc}
\eta_{\rm dc}\approx\frac{\Delta\nu}{\tilde{\gamma}_e {\cal C}\sqrt{I_{\rm PL}}},
\end{eqnarray}
where $\Delta\nu$ is the ODMR linewidth, $\tilde{\gamma}_e\approx2.8$~MHz/G is the electron gyromagnetic ratio, ${\cal C}$ is the ODMR contrast, and $I_{\rm PL}$ is the photoluminescence (PL) detection rate under continuous wave (CW) excitation. For the measurement of ac \cite{Taylor2008} and randomly fluctuating magnetic fields \cite{Cole2009,Hall2009}, the sensitivity scales as
\begin{eqnarray} \label{Eq:eta_ac}
\eta_{\rm ac}\approx\frac{1}{\tilde{\gamma}_e {\cal C}\sqrt{I_{\rm PL}t_{\rm ro}T_2}},
\end{eqnarray}
where $t_{\rm ro}$ is the readout pulse duration and $T_2$ is the spin coherence time under the employed measurement sequence (e.g., Hahn echo) \cite{Rondin2014}. The aim of this work is to investigate how the dc and ac sensitivities depend on the implantation and sample processing conditions. The role of implantation energy and dose was first analysed by comparing all samples after annealing at $950^\circ$C and acid cleaning. We then investigated the effect of a second annealing step at temperatures above $1100^\circ$C and/or an oxygen annealing step at $465^\circ$C. Double electron-electron resonance (DEER) \cite{Grotz2011,Mamin2012} as well as cross-relaxation EPR spectroscopy \cite{Hall2016,Wood2016} were performed on the NV centres and used to characterise paramagnetic defects associated with the various stages of treatment. 

\section{Methods}

\subsection{Diamond samples} \label{sec:MethodSample}

The diamond samples used in this study were made from 4 mm $\times$ 4 mm single-crystal diamond plates grown by CVD, with a thickness ranging from 30-120~$\mu$m, a bulk nitrogen content [N]~$<1$~ppb and a (100)-oriented top surface polished with a best surface roughness $<5$~nm Ra (purchased from Delaware Diamond Knives). Some of the plates were  homoepitaxially overgrown with 2~$\mu$m of high purity ([N]~$<1$~ppb) CVD diamond using $^{12}$C-enriched (99.95\%) methane. The surfaces of the overgrown and as-received samples will be referred to as as-grown (A) and polished (P) surfaces, respectively. All the plates were then laser cut into 2 mm $\times$ 2 mm plates and acid cleaned (15 minutes in a boiling mixture of sulphuric acid and sodium nitrate). The diamonds were then implanted with $^{14}$N$^+$ or $^{15}$N$^+$ ions (InnovIon) at various energies and doses, with a tilt angle of 7$^\circ$. Following implantation, the samples were annealed in a vacuum of $\sim10^{-5}$~Torr either at 950$^\circ$C for 4h (hereafter referred to as 950$^\circ$C annealing), and/or at higher temperature (hereafter referred to as 1100$^\circ$C annealing or 1200$^\circ$C annealing) using the following sequence in a similar vacuum: 6h at 400$^\circ$C, 2h ramp to 800$^\circ$C, 6h at 800$^\circ$C, 2h ramp to 1100$^\circ$C or 1200$^\circ$C, 2h at 1100$^\circ$C or 1200$^\circ$C, 2h ramp to room temperature. To remove the graphitic layer formed during the annealing at the elevated temperatures, the samples were acid cleaned using the process outlined previously. Some of the samples were further annealed at 465$^\circ$C for 4h at atmospheric pressure, a process hereafter referred to as oxygen annealing. Table \ref{Table:samples} lists the implantation parameters and processing details for each sample analysed in this work, and defines the naming convention used throughout the paper.

\begin{table*}[hbt!]
\begin{tabular}{|c|c|c|c|c|c|c|c|}
\hline
Sample & Surface & Implanted & Energy & Dose & Initial & Second & \\ 
name &  type & isotope & (keV) & (ions/cm$^2$) & annealing & annealing & Figures \\
\hline
P-6-low  & Polished & $^{15}$N & 6 & $5\times10^{11}$ & 950$^\circ$C & & 1 \\ 
P-6-high-a  & Polished &  $^{15}$N & 6 & $1\times10^{13}$ & 1200$^\circ$C & & 6 \\ 
P-6-high-b  & Polished &  $^{15}$N & 6 & $1\times10^{13}$ & 950$^\circ$C & & 2, 3, 6 \\ 
P-6-high-c  & Polished &  $^{15}$N & 6 & $1\times10^{13}$ & 950$^\circ$C & 1100$^\circ$C & 1, 2, 3, 4, 6 \\ 
P-10-low  & Polished &  $^{15}$N & 10 & $1\times10^{12}$ & 950$^\circ$C & & 3 \\
P-10-high  & Polished &  $^{15}$N & 10 & $1\times10^{13}$ & 950$^\circ$C & 1200$^\circ$C & 2, 3, 4 \\ 
P-14-high  & Polished &  $^{15}$N & 14 & $1\times10^{13}$ & 950$^\circ$C & 1200$^\circ$C & 1, 2, 3, 4 \\ 
P-30-high  & Polished &  $^{15}$N & 30 & $3\times10^{12}$ & 950$^\circ$C & 1200$^\circ$ & 3, 4 \\
A-4-low  & As grown &  $^{15}$N & 4 & $5\times10^{11}$ & 950$^\circ$C & & 1, 6 \\  								% C13 contamination
A-4-high-a  & As grown  & $^{14}$N & 4 & $1\times10^{13}$ & 950$^\circ$C & 1200$^\circ$C & 2, 3, 4, 7 \\ 
A-4-high-b  & As grown  & $^{15}$N & 4 & $1\times10^{13}$ & 950$^\circ$C & & 2, 3, 6, 7 \\ 
A-5-high-a  & As grown  & $^{14}$N & 5 & $5\times10^{12}$ & 1200$^\circ$C & & 7 \\ 
A-5-high-b  & As grown  & $^{15}$N & 5 & $1\times10^{13}$ & 950$^\circ$C & 1100$^\circ$C & 2, 3, 4, 7 \\ 
A-6-low-a  & As grown  & $^{15}$N & 6 & $5\times10^{11}$ & 950$^\circ$C & 1100$^\circ$C & 1, 4, 6 \\ 			% C13 contamination
A-6-low-b  & As grown  & $^{15}$N & 6 & $1\times10^{12}$ & 950$^\circ$C & & 3 \\								% C13 contamination
A-6-high-a  & As grown  & $^{14}$N & 6 & $1\times10^{13}$ & 1200$^\circ$C & & 6, 7 \\ 
A-6-high-b  & As grown  & $^{14}$N & 6 & $1\times10^{13}$ & 1200$^\circ$C & & 7\\ 
A-6-high-c  & As grown  & $^{15}$N & 6 & $1\times10^{13}$ & 950$^\circ$C & 1200$^\circ$C & 1, 3, 4, 5, 7 \\  
A-14-high  & As grown  & $^{15}$N & 14 & $5\times10^{12}$ & 950$^\circ$C & 1100$^\circ$C & 3, 4, 5 \\ 			% C13 contamination
\hline
\end{tabular}
\caption{Sample details. The samples are named following the convention $X$-$Y$-$Z$-$i$ where $X$ = `P' (polished surface) or `A' (as-grown surface), $Y$ is the implantation energy (in keV), $Z$ = `high' (dose $\geq3\times10^{12}$~ions/cm$^2$) or `low' (dose $\leq10^{12}$~ions/cm$^2$), and $i$ is a letter added when further distinction is needed. The last column lists the figures in which each sample appears.}	
\label{Table:samples}	
\end{table*}

\subsection{NV measurements} \label{sec:MethodNV}

The spin properties of the NV layers were investigated using a purpose-built wide-field microscope similar to that described in Refs. \cite{Simpson2016,Tetienne2017}. The diamond chips were mounted on a glass cover slip, with the NV layer facing up. The cover slips were patterned with a microwave resonator and mounted on a printed circuit board. The NV centres were excited and imaged via an inverted optical microscope with an oil-immersion 40x objective (NA~$=1.3$). The power of the 532 nm excitation laser was 300 mW at the sample with a Gaussian beam diameter of $\approx100~\mu$m. The emitted red PL was filtered (wavelength 660-735 nm) and imaged with a scientific complementary metal oxide semiconductor (sCMOS) camera providing a maximum field of view of 200~$\mu$m~$\times$ 200~$\mu$m. All measurements were performed at room temperature under ambient conditions. 

\paragraph*{Photoluminescence (PL) rate.} The PL rate, $I_{\rm PL}$, was determined by measuring the photon count rate (in million counts per second, Mcount/s) from a single 400 nm $\times$ 400 nm pixel under CW laser excitation, in zero magnetic field. The pixel size was chosen to approximately match the spatial resolution of the microscope. The signal was averaged over a 50~$\mu$m~$\times$~50~$\mu$m field of view, over which the laser intensity was roughly uniform and an order of magnitude below saturation of the NV optical cycling.  

\paragraph*{ODMR linewidth.} To determine the minimum linewidth, $\Delta\nu$, pulsed ODMR spectra were recorded at a magnetic field of $B_0=46$~G produced by a permanent magnet, aligned with a $\langle111\rangle$ crystal direction. A microwave $\pi$-pulse duration of $t_\pi=1.5~\mu$s was used, which was found to be sufficiently long to eliminate power broadening in all samples investigated \cite{Dreau2011}. We also verified that further reduction of the field of view (from 50~$\mu$m~$\times$~50~$\mu$m to 12~$\mu$m~$\times$~12~$\mu$m) did not reduce the ODMR linewidth, indicating that the linewidth is not limited by inhomogeneities in the static ($B_0$) or microwave field over the chosen field of view. Thus, the linewidth observed in the recorded spectra was representative of the $T_2^*$-limited linewidth, where $T_2^*$ is the spin dephasing time. The spectra were fitted with a sum of two or three Lorentzian peaks (accounting for the hyperfine structure from the $^{15}$N or $^{14}$N isotope, respectively) with a common linewidth, $\Delta\nu$, defined as the full width at half maximum (FWHM).

\paragraph*{Rabi contrast.} The spin readout contrast, ${\cal C}$, was determined by recording Rabi oscillations at a magnetic field of 480 G, by driving the NV transitions at a microwave frequency of $\approx1520$ MHz. At such a field, the NV nuclear spin is completely polarised, thus avoiding beating effects due to the hyperfine structure. The microwave power was adjusted to obtain a Rabi frequency of $\approx12$~MHz (i.e., $t_\pi\approx40$~ns). The curves were fitted with a damped oscillation, and ${\cal C}$ was defined as the maximum peak-to-peak contrast of the fit.  

\paragraph*{Spin coherence time.} Hahn echo decoherence curves were recorded at 480 G using a Rabi frequency of $\approx12$~MHz. At this field, several collapses and revivals caused by the $^{13}$C bath (for samples with natural isotopic concentration) are visible within the overall coherence time. The curves were fitted with an oscillation enveloped by an exponential decay, $e^{-\tau/T_2}$, where $\tau$ is the total evolution time and $T_2$ defines the Hahn echo spin coherence time. 

\paragraph*{DEER spectroscopy.} DEER spectra were recorded at a field $B_0=480$~G using a Hahn echo sequence on the NV spins with a total evolution time of 2~$\mu$s. The duration of the microwave $\pi$-pulse applied to the dark spins was adjusted to 240 ns (except in Fig. \ref{Fig5}b where it was 100 ns), resulting in a line broadening of $\sim4$~MHz, which was of the same order as the broadening induced by the inhomogeneity in $B_0$ near 480~G over the 50~$\mu$m~$\times$~50~$\mu$m field of view. With these parameters kept constant, the driving frequency, $\omega$, was swept over a 200 MHz window (1 GHz in Fig. \ref{Fig5}b) centred around the free-electron Larmor frequency $\omega_e=\gamma_eB_0\approx1350$~MHz, hereafter referred to as the $g=2$ line ($g$ is the Land{\'e} factor). The signal was normalised by alternating the termination pulse of the Hahn echo sequence with a $\frac{3\pi}{2}$ instead of a $\frac{\pi}{2}$ microwave pulse. The spectra are plotted as a function of the difference $\Delta\omega=\omega-\omega_e$.   

\paragraph*{$T_1$-EPR spectroscopy.} Cross-relaxation EPR (abbreviated as $T_1$-EPR) spectra were obtained by measuring the longitudinal relaxation rate ($\frac{1}{T_1}$) of the NV spins while scanning the magnetic field in the range $\approx400$-600 G using a permanent magnet mounted on a translation stage \cite{Wood2016,Simpson2017}. For each magnet position, an ODMR spectrum was recorded to determine the magnetic field, $B$, Rabi oscillations were recorded to determine the optimal microwave $\pi$-pulse duration and the spin relaxation curve was then measured. The PL signal for the $T_1$ measurement was normalised by a microwave $\pi$-pulse on the NV transition before readout. The resulting curves were fitted by a stretched exponential $e^{-(\tau/T_1)^n}$, where the exponent $n$ is found to be $n\approx0.7$ \cite{Simpson2017}, $\tau$ is the total evolution time, and $\Gamma_1=\frac{1}{T_1}$ defines the spin relaxation rate plotted in the $T_1$-EPR spectra. To allow direct comparison with the DEER spectra, the $T_1$-EPR spectra are plotted as a function of $\Delta\omega=2\tilde{\gamma_e}(B-B_e)$ where $B_e\approx512$~G is the resonant magnetic field for a free electron ($g=2$). The factor 2 accounts for the fact that increasing $B$ decreases the NV probe frequency and increases the frequency of the target spins at the same time \cite{Hall2016,Wood2016}.    

\paragraph*{Proton NMR measurements.} To estimate the average depth of the NV layer, NMR proton signals were detected and analysed following the method of Ref. \cite{Pham2015}. To this end, the samples were covered with immersion oil (Olympus, Type-F) and NMR spectra were acquired at a field of $\approx200$ G using the XY8 dynamical decoupling sequence with 64 or 128 microwave $\pi$-pulses, with $t_\pi\approx40$ ~ns. The spectra were fitted by including a finite proton dephasing time as a free parameter \cite{Pham2015}. The model assumes an infinitesimally thin  layer of NV centres at a depth $d$ from the proton sample, which may include protons intrinsic to the surface \cite{DeVience2015,Staudacher2015}. The proton density in the sample was taken to be $55\pm5$~nm$^{-3}$, which is a typical range for organic materials. Note that only samples enriched in $^{12}$C were measured to avoid spurious harmonics from $^{13}$C spins \cite{Loretz2015}. Moreover, to verify the nature of the signal, correlation spectra were recorded using an XY8-64 sequence \cite{Laraoui2013,Staudacher2015}. 

\paragraph*{Magnetic sensitivity.} The sensitivities $\eta_{\rm dc}$ and $\eta_{\rm ac}$ were calculated using Eqs. (\ref{Eq:eta_dc}) and (\ref{Eq:eta_ac}), respectively, using the values of $I_{\rm PL}$, $\Delta\nu$, ${\cal C}$ and $T_2$ obtained as explained above. It should be noted that Eq. (\ref{Eq:eta_dc}) corresponds to the magnetic sensitivity for CW ODMR. In pulsed ODMR, the average PL rate during the measurement is only a fraction of the CW rate, corresponding to the duty cycle of the readout laser pulse, of approximately $\frac{t_{\rm ro}}{t_{\rm ro}+t_\pi}$. When $t_{\rm ro}\ll t_\pi$, an expression for the optimal sensitivity can be derived, with an optimised $\pi$-pulse duration $t_\pi\approx T_2^*$ \cite{Dreau2011}. However, in our experiments the readout laser pulse is relatively long ($t_{\rm ro}=5~\mu$s, commensurate with the low laser power density used), resulting in a duty cycle close to unity. In this situation, Eq. (\ref{Eq:eta_dc}) provides a good estimate of the sensitivity that would be obtained in a pulsed ODMR experiment using $t_\pi\approx T_2^*$, for which $\Delta\nu$ is approaching the minimum linewidth imposed by spin dephasing, and ${\cal C}$ is only marginally reduced relative to the maximum contrast observed in Rabi oscillations \cite{Dreau2011}. Note that an additional factor $\sim3$ could be gained in sensitivity using saturation laser power. Likewise, Eq. (\ref{Eq:eta_ac}) was derived under the assumption $t_{\rm ro}\ll T_2$ \cite{Taylor2008} but nevertheless provides a good estimate of the sensitivity that would be obtained in a Hahn echo measurement with a total evolution time $\tau\approx T_2/2$. 

\begin{figure*}[hbt!]
	\begin{center}
		\includegraphics[width=0.85\textwidth]{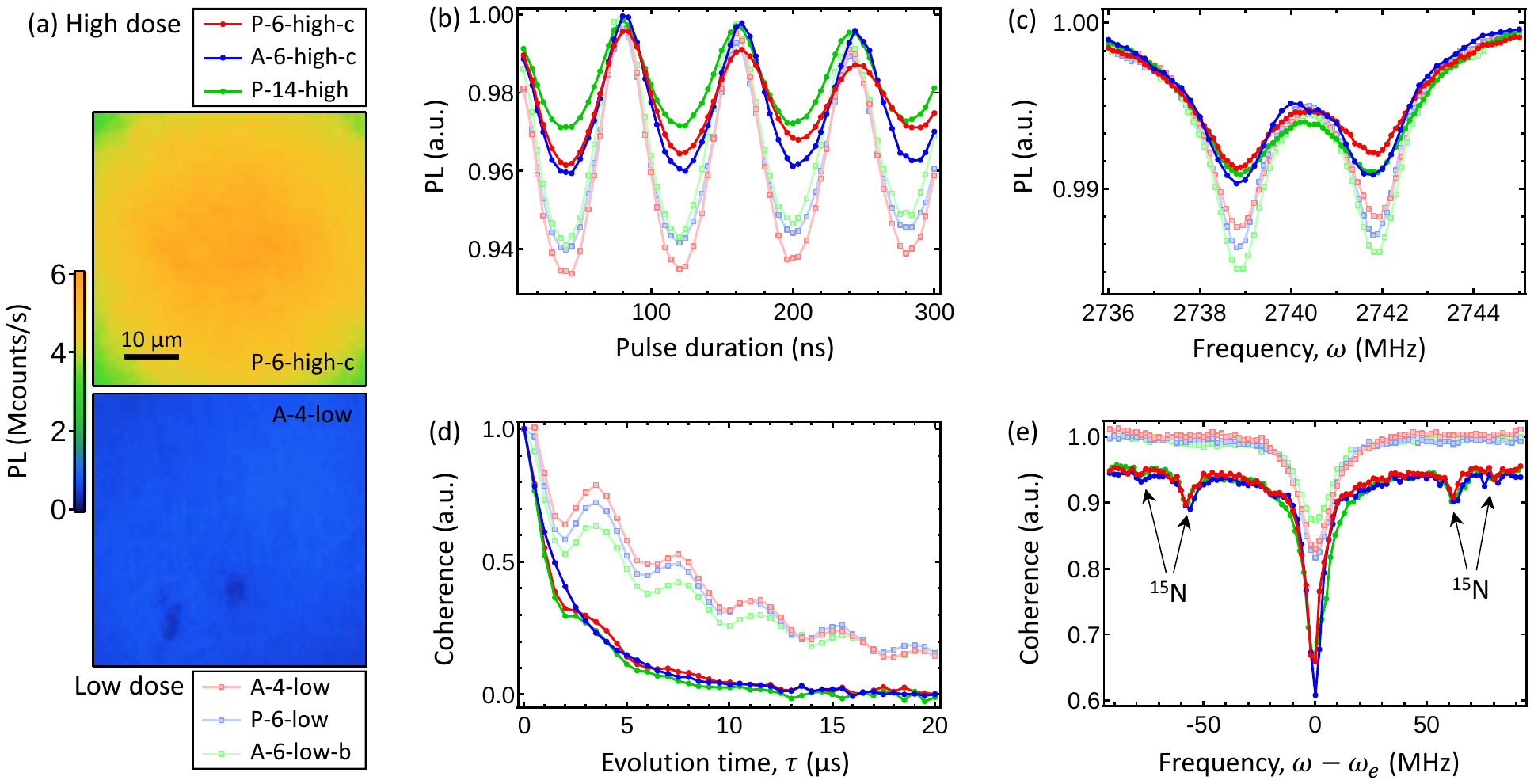}
		\caption{(a) List of selected samples investigated in (b-e) implanted with high dose (top box) and low dose (bottom box), and PL snapshots of two representative samples (top image, dose $10^{13}$~ions/cm$^2$; bottom image, dose $5\times10^{11}$~ions/cm$^2$). The total laser power is 300 mW, with a laser spot diameter of $\approx100~\mu$m. In all subsequent measurements, the PL signal is integrated over the whole $50~\mu$m~$\times~50~\mu$m area. (b) Rabi oscillations recorded at 480 G for the 6 samples listed in (a). (c) ODMR spectra recorded at 46 G. (d) Hahn echo decoherence curves recorded at 480 G. (e) DEER spectra recorded at 480 G, with the high dose curves vertically offset for clarity. The four lines associated with the hyperfine structure of substitutional $^{15}$N impurities (P1 defects) are indicated by arrows. The details of the measurements in (b-e) are given in Sec. \ref{sec:MethodNV}.}
		\label{Fig1}
	\end{center}
\end{figure*} 

\section{Results}

\subsection{Effect of implantation parameters}

We begin by examining the effect of implantation parameters (energy and dose) on the spin properties of the formed NV layers. To this end, we prepared a total of 14 diamond samples that were laser cut and acid cleaned prior to implantation, implanted with nitrogen ions at various energies and doses, annealed at 950$^\circ$C and acid cleaned (details are given in Sec. \ref{sec:MethodSample}). Each sample was then measured to obtain the various parameters governing the magnetic sensitivities expressed by Eqs. (\ref{Eq:eta_dc}) and (\ref{Eq:eta_ac}). 

To allow for a first qualitative assessment of the NV properties, we show data from a few selected samples (Fig. \ref{Fig1}). Fig. \ref{Fig1}a shows PL images of two different samples under CW excitation, as obtained directly on the sCMOS camera. The main difference between the two images is the intensity, i.e. the PL rate, which is significantly larger for the sample with the largest implantation dose ($10^{13}$~ions/cm$^2$ against $5\times10^{11}$~ions/cm$^2$). This is expected, as the dose impacts directly on the number of NV centres that are formed, hence on the PL rate per unit surface area. Less trivial is the effect of implantation dose on the Rabi contrast (${\cal C}$, Fig. \ref{Fig1}b), the ODMR linewidth ($\Delta\nu$, Fig. \ref{Fig1}c) and the spin coherence time ($T_2$, Fig. \ref{Fig1}d). The data shown suggests a clear trend whereby ${\cal C}$ is larger, $\Delta\nu$ is narrower and $T_2$ is longer for lower doses ($\leq10^{12}$~ions/cm$^2$) compared to high doses ($10^{13}$~ions/cm$^2$). This trend seems to be relatively independent of the implantation energy, for instance, $T_2$ is similar for 6 and 14 keV samples at the highest dose ($T_2=2.5-3~\mu$s) and significantly shorter than at the lowest dose ($T_2=10-12~\mu$s), suggesting that these quantities are dominated by implantation-induced defects with negligible contributions from the surface proximity effects seen to limit $T_2$ times in single-NV samples \cite{Rosskopf2014,Myers2014,Romach2015}. Fig. \ref{Fig1}e shows DEER spectra of the same samples. For the highest dose, the hyperfine structure characteristic of the substitutional $^{15}$N impurities (known as P1 defects \cite{Smith1959}) is clearly visible. In addition, all spectra feature a strong $g=2$ peak whose amplitude scales with the dose and not the energy. This is a direct indication that the implantation process creates a significant number of paramagnetic defects, which in turn may limit the spin coherence of the NV layer, as will be discussed further later.

\begin{figure}[htb]
	\begin{center}
		\includegraphics[width=0.49\textwidth]{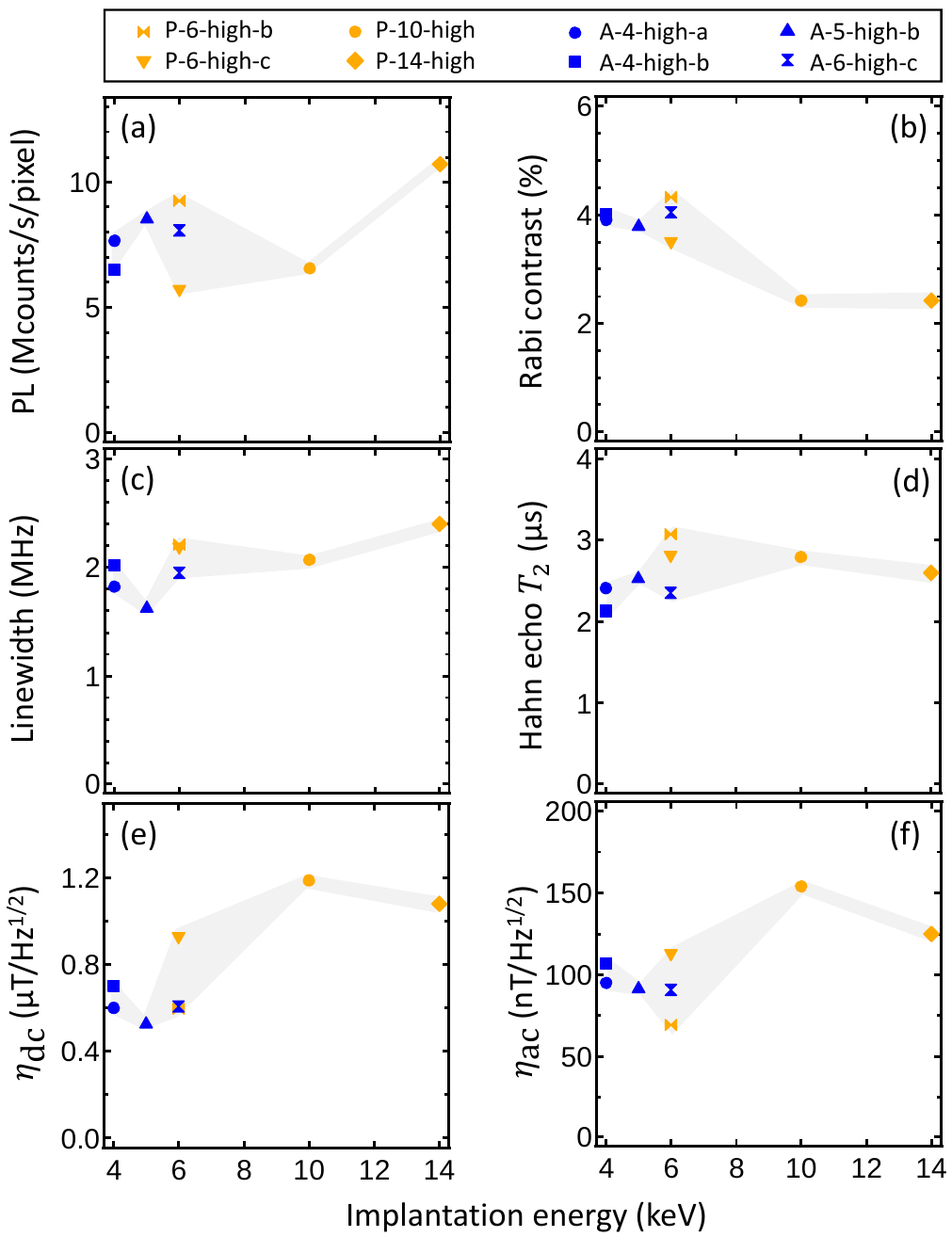}
		\caption{(a-d) PL rate (a), Rabi contrast (b), ODMR linewidth (c) and Hahn echo $T_2$ time (d) as a function of implantation energy for 7 different samples implanted at a dose of $10^{13}$~ions/cm$^2$ and annealed at 950$^\circ$C. For each sample, the values are averaged over a $50~\mu$m~$\times~50~\mu$m area. The vertical error bars (one standard deviation) are smaller than the symbols hence not shown. The grey shaded areas are a guide to the eye. (e,f) Theoretical sensitivity to dc (e) and ac (f) magnetic fields calculated from Eqs. (\ref{Eq:eta_dc}) and (\ref{Eq:eta_ac}) using the parameters measured in (a-d).}
		\label{Fig2}
	\end{center}
\end{figure}

\begin{figure}[htb]
	\begin{center}
		\includegraphics[width=0.49\textwidth]{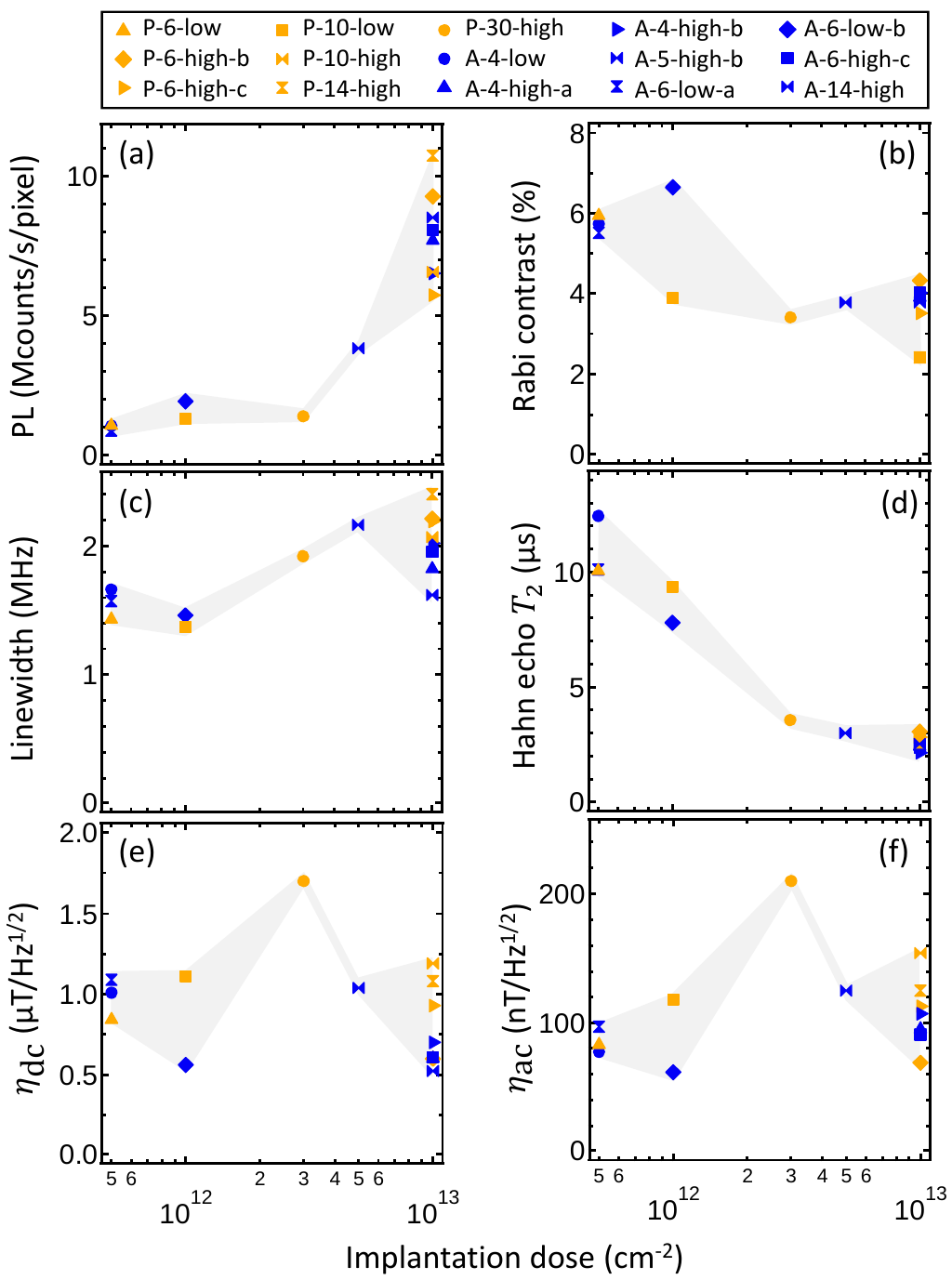}
		\caption{(a-d) PL rate (a), Rabi contrast (b), ODMR linewidth (c) and Hahn echo $T_2$ time (d) as a function of implantation dose for 14 different samples implanted at various energies and annealed at 950$^\circ$C (see sample details in Table \ref{Table:samples}). The vertical error bars (one standard deviation) are smaller than the symbols hence not shown. The grey shaded areas are a guide to the eye. (e,f) Theoretical sensitivity to dc (e) and ac (f) magnetic fields deduced from (a-d).}
		\label{Fig3}
	\end{center}
\end{figure}

The effects of implantation energy and dose are further explored in Figs. \ref{Fig2} and \ref{Fig3}, using the parameters extracted from complete sample sets. Figs. \ref{Fig2}a-d show the average PL rate per pixel ($I_{\rm PL}$), Rabi contrast (${\cal C}$), ODMR linewidth ($\Delta\nu$) and coherence time ($T_2$), respectively, as a function of energy for 7 samples implanted at the highest dose ($10^{13}$~ions/cm$^2$). There is no obvious variation of these parameters with energy, considering the relatively large sample-to-sample variations observed for a given energy, except for an apparent decrease in the Rabi contrast with increasing energy. As figures of merit, we calculated the dc and ac sensitivities from Eqs. (\ref{Eq:eta_dc}) and (\ref{Eq:eta_ac}) using the measured parameters (Figs. \ref{Fig2}e,f). For energies $\leq6$~keV, we obtain an average of $\approx0.7~\mu$T/Hz$^{1/2}$ and $\approx90$~nT/Hz$^{1/2}$, respectively, from a single 400~nm~$\times$~400~nm pixel, and slightly larger values at higher energy.

As anticipated from Fig. \ref{Fig1}, the implantation dose is a key parameter, as shown in Figs. \ref{Fig3}a-d. Despite significant sample-to-sample variations, clear trends are observed for all parameters measured. The PL rate (Fig. \ref{Fig3}a) increases with the dose, but not linearly: $I_{\rm PL}$ increases by, at most, a factor 2 from the lowest dose of $5\times10^{11}$~ions/cm$^2$ up to $3\times10^{12}$~ions/cm$^2$, and by a factor 6-11 from the lowest dose to the highest dose of $10^{13}$~ions/cm$^2$. This suggests that the N-to-NV$^-$ conversion rate decreases when increasing the dose \cite{Pezzagna2010}, where NV$^-$ denotes the negatively charged state of the NV centre, which is the dominant contribution to the detected PL in our measurements. The Rabi contrast (Fig. \ref{Fig3}b) decreases by a factor 2 on average from lowest to highest dose. This is possibly related to NV charge state dynamics \cite{Waldherr2011,Aslam2013a,Hopper2017} or to variations in the radiative and nonradiative transition rates of the NV$^-$ charge state alone \cite{Bogdanov2017}, which may both be affected by the local density of defects hence the dose. The ODMR linewidth exhibits a more modest change with dose, increasing from 1.5 to 2 MHz on average for lowest to highest dose, with a relatively large variability between samples (Fig. \ref{Fig3}c). Comparatively, the $T_2$ time exhibits smaller sample-to-sample variations and a larger dose dependence, ranging from 10-12~$\mu$s at the lowest dose and 2-3~$\mu$s at the highest dose. Both the linewidth and $T_2$ time are related to the magnetic noise generated by fluctuating spins in the diamond lattice or on the diamond surface \cite{Hall2014,Yang2017}. Our observations thus suggest that the noise in the frequency range $10^4-10^6$ Hz (which governs $T_2$) is dominated by implantation-induced paramagnetic defects within the lattice. This is further illustrated by noting that sample P-30-high, which was implanted at 30 keV (the largest energy of this study, corresponding to NV centres about 70 nm deep on average \cite{Toyli2010}), shows no improvement in $T_2$ relative to much shallower NV samples.     

\begin{figure*}[htb!]
	\begin{center}
		\includegraphics[width=0.95\textwidth]{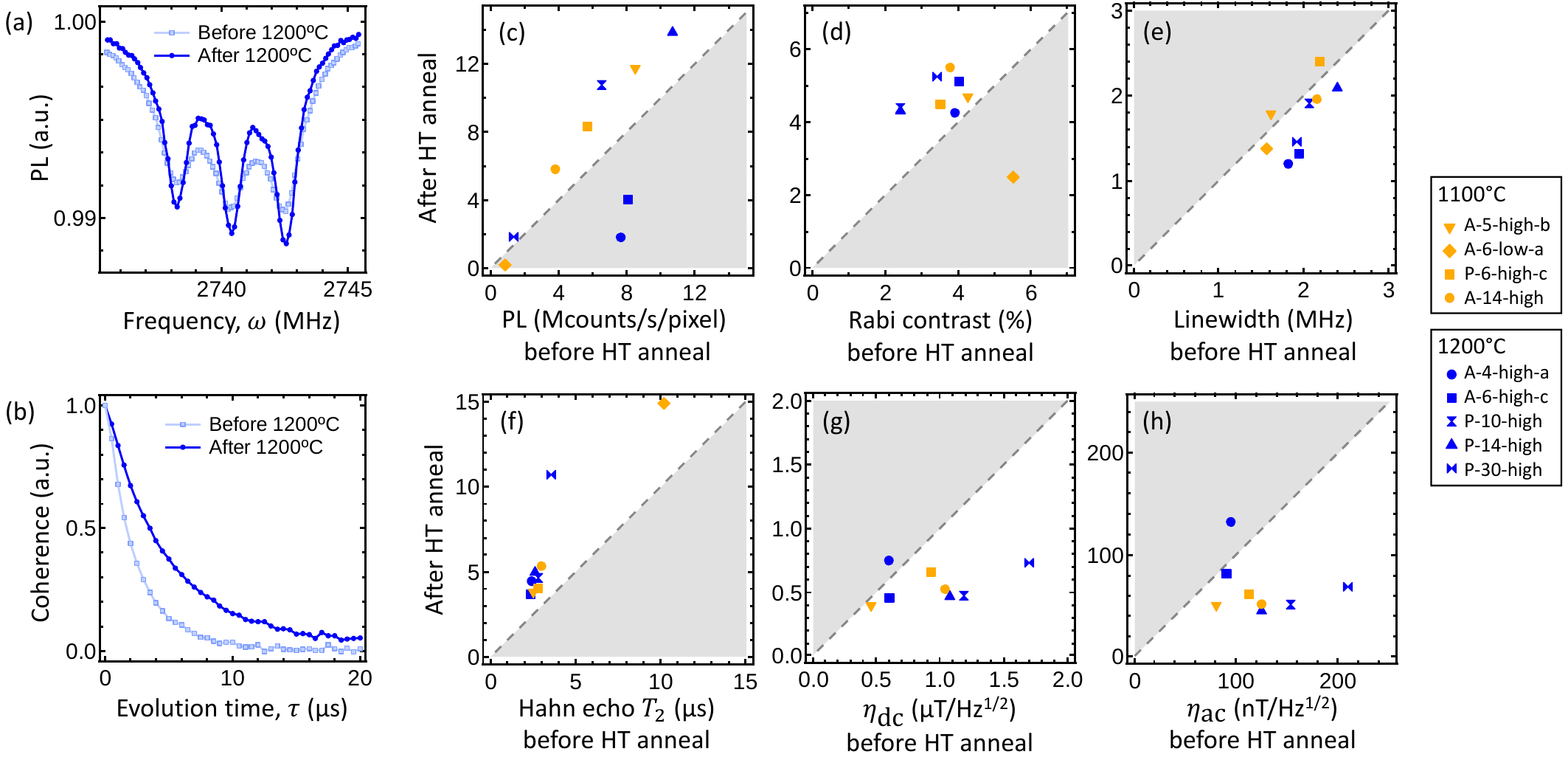}
		\caption{(a,b) ODMR spectra (a) and Hahn echo decoherence curves (b) recorded for sample A-4-high-a before and after a second annealing step at 1200$^\circ$C. (c-f) PL rate (c), Rabi contrast (d), ODMR linewidth (e) and Hahn echo $T_2$ time (f) after a second annealing step at a higher temperature (HT, 1100$^\circ$C or 1200$^\circ$C) plotted as a function of the value measured before the HT annealing, for 9 different samples (see sample details in Table \ref{Table:samples}). The error bars (one standard deviation) are smaller than the symbols hence not shown. (g,h) Theoretical sensitivity to dc (g) and ac (h) magnetic fields deduced from (c-f). Note: in (g,h), the data point for sample A-6-low-a lies outside the graph (poor sensitivity due to very low PL rate after HT annealing). In (c-h), the grey dashed line corresponds to no change in the plotted quantity upon HT annealing; the grey shaded area corresponds to a deterioration in the plotted quantity, i.e. decreased PL rate, Rabi contrast and $T_2$, and increased linewidth and sensitivity; the non-shaded area in all plots therefore corresponds to an improvement.}
		\label{Fig4}
	\end{center}
\end{figure*}

In summary, a larger dose is advantageous as it provides an increased PL signal, but this is at the expense of a reduced contrast and coherence time, and increased linewidth. Overall, however, the magnetic sensitivity shows no obvious correlation with the dose, spreading over the range 0.5-1.7~$\mu$T/Hz$^{1/2}$ for dc fields (Fig. \ref{Fig3}e), and 50-220~nT/Hz$^{1/2}$ for ac fields (Fig. \ref{Fig3}f). Thus, for the samples considered here, there seems to be no optimal dose/energy combination as far as the magnetic sensitivity is concerned. It is important to note that the sensitivity was calculated assuming photon shot noise to be the dominant source of noise. However, the sensitivity may be deteriorated at low PL intensity due to other sources of noise such as readout noise and dark counts, both of which are present in a sCMOS camera such as the one used in this work. This is particularly relevant for long sensing sequences (i.e. with a laser duty cycle of $10^{-2}$ or less) based on dynamical decoupling or spin relaxation ($T_1$) measurements, for which dark counts can be of comparable magnitude as the PL signal in the low NV density samples (i.e., dose $\lesssim10^{12}$~ions/cm$^2$). Therefore, higher dose samples (typically, $10^{13}$~ions/cm$^2$) are generally preferred over low dose, while using a low energy (e.g., 4 keV) maximises the signal from a given external sample without affecting the sensitivity. Finally, we note that the type of surface (polished or as grown) seems to have no impact on the measured properties in these samples, whereas polishing damage is typically detrimental for single NV centres \cite{FavarodeOliveira2015}, although further work is needed to investigate the impact of polishing damage in our overgrown layers as they are only 2 $\mu$m thick.

\subsection{Effect of annealing temperature}

Having analysed the role of implantation parameters, we now move on to investigate the effect of the post-implantation processing conditions, namely the annealing temperature and the subsequent surface treatment. The annealing temperature must be chosen to allow migration of vacancies to form stable NV defects, with 800$^\circ$C and 950$^\circ$C being the most commonly used temperatures. On the other hand, the exact temperature also affects how other defects form or anneal out, which has consequences for the charge and spin environment around the NV centres. In particular, it has been found through bulk EPR studies that the number of paramagnetic defects is dramatically reduced for annealing temperatures in the range 1100$^\circ$C-1200$^\circ$C \cite{Lomer1973,Yamamoto2013}. Furthermore, it was shown that for diamond samples implanted with nitrogen ions at low dose ($<10^9$~ions/cm$^2$) and relatively high energy ($>100$~keV), annealing at 1200$^\circ$C had a beneficial impact on the NV spin coherence time \cite{Yamamoto2013,Naydenov2010}.  However, the effect of annealing temperature for dense layers of near-surface NV centres has not been studied so far, despite the expected increase in the density of relevant defects with these implants. 

We used 9 samples initially annealed at 950$^\circ$C, and annealed them a second time at a higher temperature, either 1100$^\circ$C or 1200$^\circ$C, followed by acid cleaning (see experimental details in Sec. \ref{sec:MethodSample}). An example of typical ODMR spectra and decoherence curves, measured for a given sample before and after a high temperature (HT) annealing, are shown in Figs. \ref{Fig4}a and \ref{Fig4}b, respectively, illustrating a clear improvement in the ODMR linewidth and $T_2$ time after the second annealing. Figs. \ref{Fig4}c-f show the PL rate, Rabi contrast, ODMR linewidth and $T_2$ time before and after the HT annealing for each of the 9 samples. The parameter value before HT annealing is shown on the $x$ axis and its value after HT annealing is shown on the $y$ axis; data on the $x=y$ line (dashed grey line) indicate no change due to HT annealing. For 6 of these samples (implantation energy 5-14 keV), the PL rate increased by a factor 1.3-1.6, which indicates an increased N-to-NV$^-$ conversion rate. Such an increase is consistent with the findings of Ref. \cite{Yamamoto2013} for deep NV centres (4-13 MeV). On the other hand, the other 3 samples (implantation energy 4-6 keV) exhibit a significantly reduced PL, by a factor 2-4. This is possibly due to etching of the diamond surface, or to some of the existing NV centres being annealed out or converted into larger defect clusters; however, this effect must be very sensitive to the exact conditions of the process (e.g., the residual gaseous species present during the HT annealing) since other nominally similar samples had their PL increased instead of decreased upon HT annealing. 

The Rabi contrast (Fig. \ref{Fig4}d) is increased for all samples, by a factor 1.1-1.8, except sample A-6-low-a which had the smallest PL rate following HT annealing and had a contrast decreased by a factor 2. Similarly, the ODMR linewidth is improved for all samples but 2, and the coherence time is increased for all samples, by a factor 1.5-3. Consequently, the magnetic sensitivity (Figs. \ref{Fig4}e and \ref{Fig4}f) is improved for most samples (7 out of 9), by a factor up to 2.5 for the dc sensitivity, and up to 3.3 for the ac sensitivity, with best values of 400~nT/Hz$^{1/2}$ and 40~nT/Hz$^{1/2}$ from a single pixel, respectively. We note that these results do not seem to depend on the exact temperature of the second annealing step, whether it is 1100$^\circ$C or 1200$^\circ$C.
 
\begin{figure}[tb!]
	\begin{center}
		\includegraphics[width=0.45\textwidth]{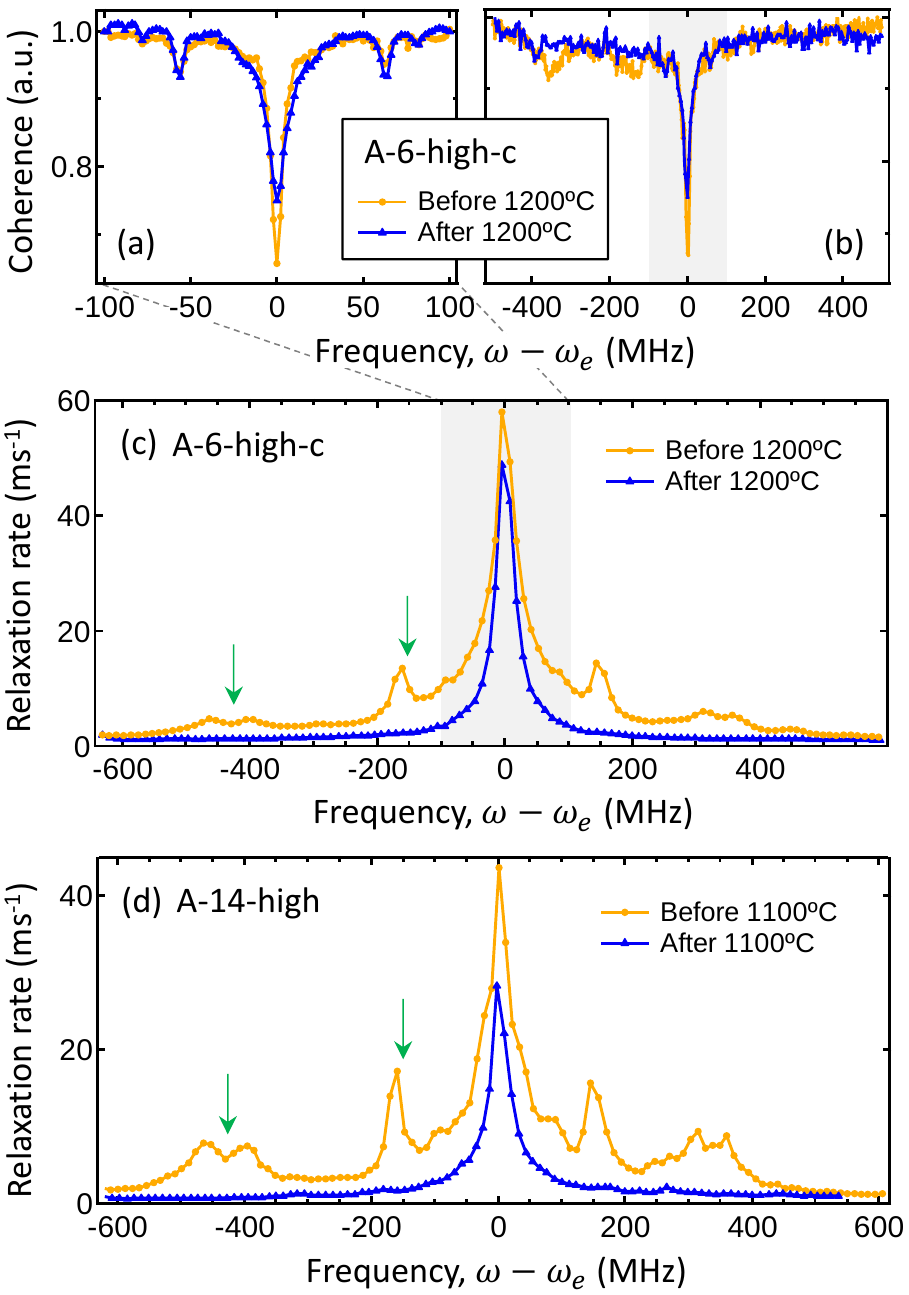}
		\caption{(a,b) DEER spectra recorded for sample A-6-high-c before (orange) and after (blue) a second annealing at 1200$^\circ$C. The hyperfine structure due to $^{15}$N is visible in (a), whereas the broader range in (b) uncovers additional features. (c) $T_1$-EPR spectra of the same sample as in (a,b). The green arrows indicate the approximate positions of the lines observed in the room temperature EPR spectrum in Ref. \cite{Twitchen1999}, ascribed to the divacancy (R4/W6) centre. The grey shaded area in (b,c) depicts the frequency range used in (a). (d) $T_1$-EPR spectra of sample A-14-high before and after a second annealing at 1100$^\circ$C.}
		\label{Fig5}
	\end{center}
\end{figure}

\begin{figure*}[htb!]
	\begin{center}
		\includegraphics[width=0.99\textwidth]{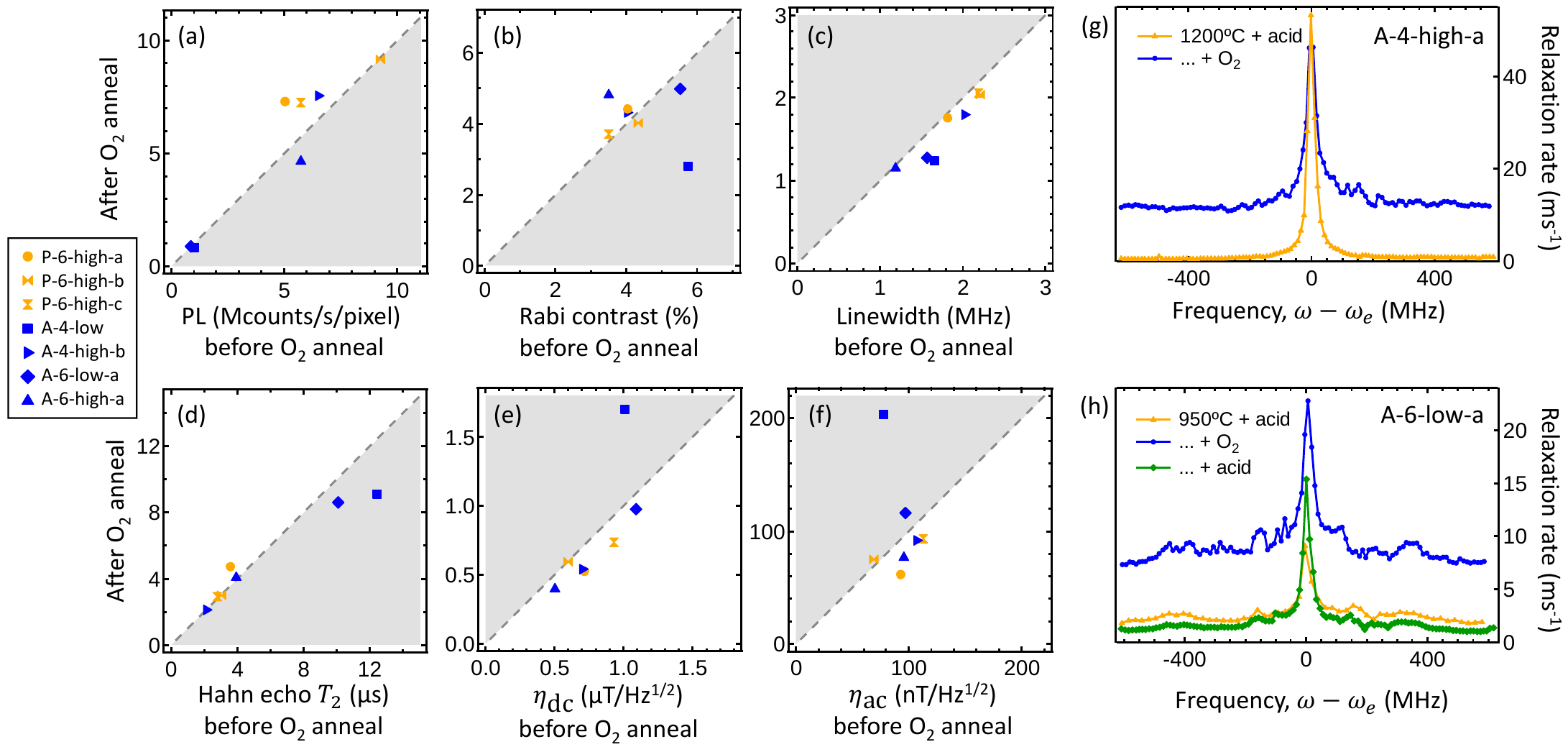}
		\caption{(a-d) PL rate (a), Rabi contrast (b), ODMR linewidth (c) and Hahn echo $T_2$ time (d) after an oxygen annealing plotted as a function of the value measured before the oxygen annealing, for 7 different samples (see sample details in Table \ref{Table:samples}). The error bars (one standard deviation) are smaller than the symbols hence not shown. (e,f) Theoretical sensitivity to dc (e) and ac (f) magnetic fields deduced from (a-d). The grey dashed line corresponds to no change in the plotted quantity upon O$_2$ annealing; the grey shaded area corresponds to a deterioration in the plotted quantity. (g,h) $T_1$-EPR spectra of samples A-4-high-a (g) and A-6-low-a (h) measured consecutively: after the initial acid cleaning (red); after an additional O$_2$ annealing (blue); and after an additional acid cleaning (green, measured for A-6-low-a only).}
		\label{Fig6}
	\end{center}
\end{figure*}

To gain more insight into the origin of these improvements upon HT annealing, we performed EPR spectroscopy using the NV layer as a probe. Namely, we used two complementary techniques, DEER spectroscopy \cite{Grotz2011,Mamin2012} and cross-relaxation spectroscopy ($T_1$-EPR) \cite{Hall2016,Wood2016}, and compared spectra recorded before and after HT annealing. Fig. \ref{Fig5}a shows DEER spectra for a diamond implanted at 6 keV with a dose of $10^{13}$~ions/cm$^2$, initially annealed at 950$^\circ$C and then further annealed at 1200$^\circ$C. The spectra reveal the hyperfine structure associated with P1 centres, which is left unchanged by the HT annealing, whereas the amplitude of the $g=2$ line is slightly reduced. DEER spectra acquired over a broader range (Fig. \ref{Fig5}b) show the presence of additional features before HT annealing, with a linewidth of 50-100~MHz indicating that the responsible spins have a relaxation time ($T_1$) of the order of 10-20~ns. Such short-lived species are easier to detect using $T_1$-EPR spectroscopy \cite{Hall2016,Wood2016}, which monitors the NV longitudinal relaxation rate $\frac{1}{T_1}$ while varying the magnetic field to map the cross-relaxation resonances between the NV spins and nearby paramagnetic defects. The resulting spectra (Fig. \ref{Fig5}c) show a $g=2$ line that is much broader than in the DEER spectra (at least by a factor 5), which implies that the noise spectrum is dominated by short-lived species to which DEER is poorly sensitive. On each side of the $g=2$ line, several features are visible in the $T_1$-EPR spectrum measured before the HT annealing, one at approximately +150/$-160$~MHz relative to the $g=2$ line, and a doublet centred at approximately +340/$-430$~MHz. We tentatively ascribe these lines to the R4/W6 centre, which is a neutral divacancy defect V$_2^0$ \cite{Twitchen1999}. The green arrows in Fig. \ref{Fig5}c indicate the positions of the two lines observed in the room temperature EPR spectrum of R4/W6 in Ref. \cite{Twitchen1999}, recorded with the field aligned along a $\langle111\rangle$ crystallographic axis similar to our measurements. These lines match well our $T_1$-EPR spectrum, apart from the splitting of the outer line which is not clearly resolved in Ref. \cite{Twitchen1999}. This apparent splitting could also arise from the presence of additional lines associated with longer multivacancy chains, such as the trivacancy defect V$^0_3$ (R5) \cite{Lomer1973,Yamamoto2013}. We note that the linewidth measured in Ref. \cite{Twitchen1999} for the R4/W6 centre at room temperature is also in good agreement with our data. The same features were observed for all the other samples we measured prior to HT annealing (another example is shown in Fig. \ref{Fig5}d), except for the lowest implantation dose where only the $g=2$ line could be clearly resolved (see Fig. \ref{Fig6}h). 
 
The divacancy centre is known to anneal out above about 1100$^\circ$C \cite{Twitchen1999}. Other defects, including V$^0_3$, anneal out at similar temperatures, typically above 1000$^\circ$C \cite{Lomer1973,Yamamoto2013,Iakoubovskii2002}, leaving mostly defects with an isotropic $g=2$ EPR signature. In our samples, we found that the side peaks in the $T_1$-EPR spectrum completely disappeared after HT annealing, either at 1200$^\circ$C (Fig. \ref{Fig5}c) or at 1100$^\circ$C (Fig. \ref{Fig5}d). In addition, the $g=2$ line became narrower, with a significantly reduced integral. This means that the overall density of residual paramagnetic defects surrounding the NV centres was significantly reduced by the HT annealing, which is believed to be the main reason why the NV properties were improved overall. It is worth mentioning that an alternative strategy to remove multivacancy defects is to prevent their formation during the initial annealing by charging the lattice, as was proposed and demonstrated in Ref. \cite{DeOliveira2017}. 

\subsection{Effect of oxygen annealing} \label{sec:O2}

As discussed previously, the properties of dense layers of NV centres are mostly limited by residual paramagnetic defects induced by the implantation, which is why they are sensitive to the annealing temperature. However, surface defects may also play a role in our samples, especially at the lowest implantation dose and/or at low energy (i.e., near-surface NV centres). The impact of surface treatment on the properties of shallow NV centres has been the subject of many studies in the regime of very low implantation dose (typically $\leq10^9$~ions/cm$^2$), for which implantation-related damage may be negligible \cite{Rosskopf2014,FavarodeOliveira2015,Kageura2017}. In this regime, one aspect to consider is the charge stability of the NV centres, which depends on the balance between the density of donors (e.g., nitrogen impurities) and the density of acceptor states on the surface. It is known that annealing the diamond at 450-550$^\circ$C in oxygen, immediately following the initial annealing or after an acid cleaning step, results in stable negatively charged NV centres \cite{Fu2010,Rondin2010,Kim2014,Yamano2017}, whereas acid cleaning alone (or even after oxygen annealing) generally results in NV centres with poor charge stability and reduced Rabi contrast \cite{Yamano2017}. Furthermore, oxygen annealing was observed to enhance the spin coherence time of very shallow NV centres \cite{Lovchinsky2016}. 

\begin{figure*}[htb!]
	\begin{center}
		\includegraphics[width=0.85\textwidth]{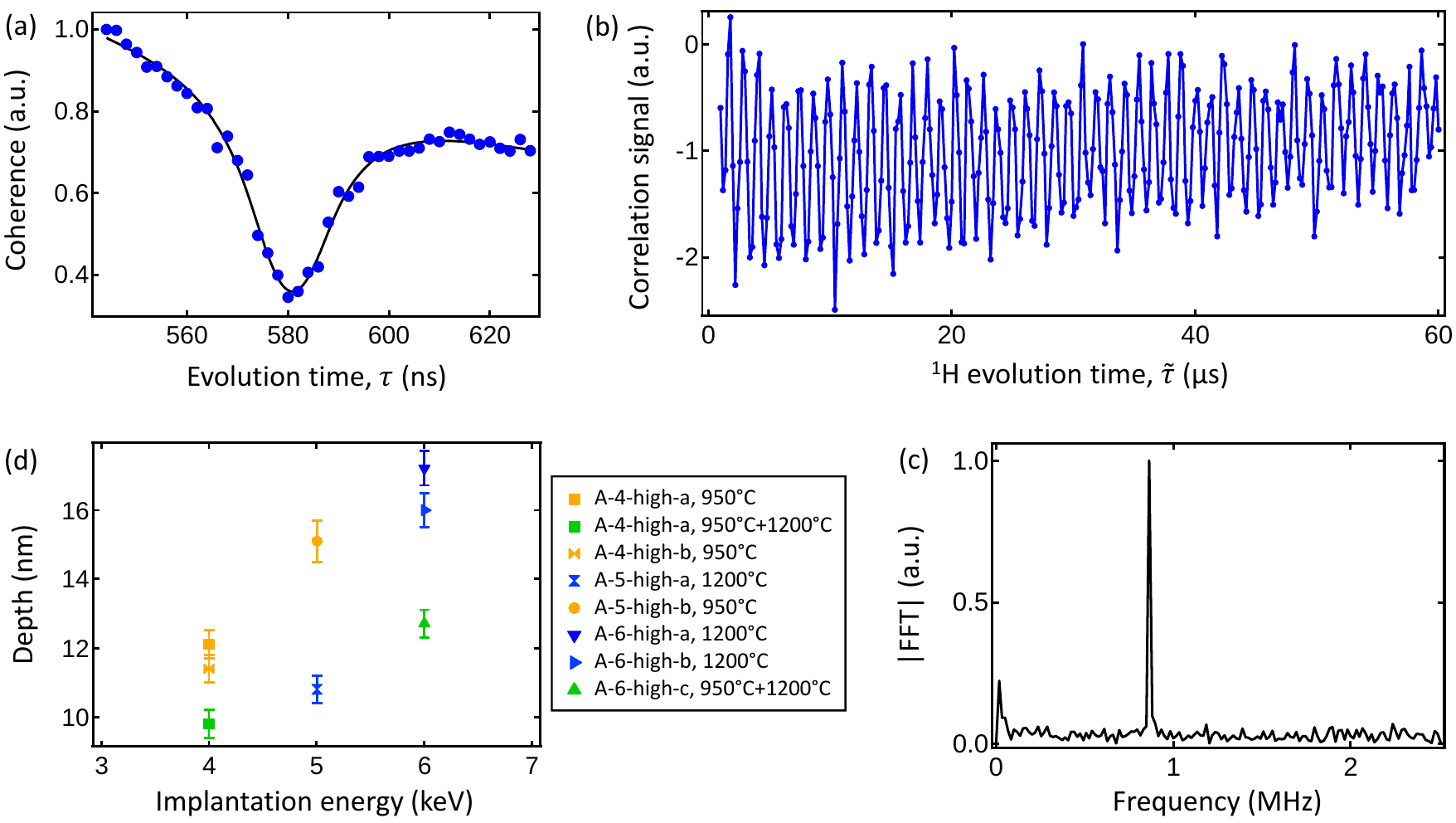}
		\caption{(a) NMR spectrum obtained on sample A-4-high-a using the XY8-128 sequence at 202 G. The dip at $\tau\approx580$~ns is the signature of $^1$H precession, where $\tau$ is the inter-pulse spacing. The solid line is a fit to the data using the same model as in Ref. \cite{Pham2015}. (b) Correlation signal obtained by fixing the inter-pulse spacing to $\tau=580$~ns and varying the time $\tilde{\tau}$ between two XY8-64 sequences. (c) Fourier transform of the correlation signal shown in (b), revealing a sharp peak at the $^1$H Larmor frequency. (d) Mean depth extracted from the XY8 spectrum plotted as a function implantation energy for 7 different samples, with sample A-4-high-a measured both before and after a 1200$^\circ$C annealing.}
		\label{Fig7}
	\end{center}
\end{figure*} 

To assess the effect of oxygen annealing on dense NV layers, we used 7 samples initially annealed in vacuum (at 950$^\circ$C or 1200$^\circ$C) and acid cleaned, and annealed them at 465$^\circ$C in oxygen for 4h. The comparative results are shown in Figs. \ref{Fig6}a-d for the PL rate (a), Rabi contrast (b), ODMR linewidth (c) and $T_2$ (d). On average, the oxygen annealing has no net effect on the PL rate and Rabi contrast, but provides a modest improvement in the linewidth, suggesting that the low-frequency magnetic noise is sensitive to the surface treatment. The $T_2$ time was found unchanged except for a decrease for samples A-4-low and A-6-low-a, which had the largest $T_2$ before annealing, due to low implantation dose ($5\times10^{11}$~ions/cm$^2$, energy 4 and 6 keV, respectively). The most dramatic effect is on sample A-4-low (shallowest NVs), for which $T_2$ was reduced from $12.5~\mu$s to $9.1~\mu$s, and the Rabi contrast from 5.7\% to 2.8\%. This indicates that in this regime of dose/energy, the surface treatment does play a role on the magnetic noise at MHz frequencies and/or on the charge stability of the NV layer. Overall, the magnetic sensitivities exhibit a modest improvement (if any) upon oxygen annealing (Figs. \ref{Fig6}e,f), except for sample A-4-low which shows a clear deterioration.

To probe the noise at GHz frequencies, we recorded $T_1$-EPR spectra before and after the oxygen annealing, with two examples shown in Figs. \ref{Fig6}g and \ref{Fig6}h. A very broad feature is consistently observed after the oxygen annealing, adding to the initial $g=2$ line (which remains about 50~MHz wide). This broad feature extends over at least 1~GHz, corresponding to a magnetic noise with correlation time $<1$~ns. Consequently, the NV relaxation time is dramatically shortened even far off resonance, e.g. $T_1\approx80~\mu$s for sample A-4-high-a (Fig. \ref{Fig6}g) and $T_1\approx120~\mu$s for A-6-low-a (Fig. \ref{Fig6}h) at a magnetic field of 400 G (i.e. $\approx600$~MHz away from the $g=2$ cross-relaxation resonance), against $T_1\approx1.8$~ms and $T_1\approx500~\mu$s before the oxygen annealing, respectively. For sample A-4-low (data not shown), $T_1$ was reduced by as much as two orders of magnitude, from $\approx1.7$~ms to $\approx18~\mu$s. Such broadband noise explains why $T_2$ was shortened in the low dose samples (see Fig. \ref{Fig6}d). The initial $T_1$-EPR spectrum could be recovered by a simple acid cleaning (green curve in Fig. \ref{Fig6}h). These observations, although not fully understood currently, suggest a complex interplay between the spin/charge dynamics of the surface states, which are clearly affected by the oxygen annealing, and the dose-dependent dynamics of the defects within the lattice. 

\subsection{Depth measurements}

Finally, we performed NMR measurements in order to estimate the average depth of the NV centres, which is a key parameter for sensing experiments. We applied immersion oil to the sample and used XY8 dynamical decoupling \cite{Staudacher2013,DeVience2015} to detect the precession of the $^1$H spins under a magnetic field of 202~G, which corresponds to a Larmor frequency of $\approx860$~kHz. An example spectrum is shown in Fig. \ref{Fig7}a, obtained for sample A-4-high-a after the HT annealing (4 keV implantation energy). The proton signal appears as a dip in the NV coherence at the expected evolution time $\tau\approx580$~ns (time between consecutive $\pi$-pulses). The nature of the signal was verified by performing correlation spectroscopy \cite{Laraoui2013,Staudacher2015} (see Fig. \ref{Fig7}b for the time domain signal and Fig. \ref{Fig7}c for the Fourier transform), revealing a sharp peak at the $^1$H Larmor frequency (860 kHz). To estimate the NV depth, the XY8 data was fitted following the procedure of Ref. \cite{Pham2015}, where we assume that all the NV centres lie at a constant depth $d$ below the diamond surface. Using a proton density of $55\pm5$~nm$^{-3}$, we find a depth $d=9.8\pm0.4$~nm for this sample. Fig. \ref{Fig7}d shows the depth as a function of the implantation energy measured for different samples from 4 to 6 keV, which ranges from $d\approx10$ to 17 nm. These values are consistent with simulations of ion implantation that take into account tilt angle and channelling effects \cite{FavarodeOliveira2015,Lehtinen2016}, which indicate an ion range between 5 and 20 nm for a 5 keV implant. We note that Fig. \ref{Fig7}d includes measurements of samples both before and after HT annealing. We were able to measure NMR spectra before HT annealing despite a much smaller absolute contrast due to much shorter coherence time. For sample A-4-high-a, in particular, the depth was found to decrease by about 2~nm upon HT annealing (1200$^\circ$C), which is attributed to graphitisation of the top 2 nm and subsequent removal by acid cleaning.

\section{Conclusion}

In summary, we investigated the spin properties of dense layers of near-surface NV centres in diamond created by nitrogen ion implantation at energies from 4-30 keV and doses from $5\times10^{11}$-$10^{13}$~ions/cm$^2$. We found that, in this regime, the spin properties (Rabi contrast, ODMR linewidth and spin coherence time) are mostly governed by the implantation dose due to the related number of paramagnetic defects, while the energy (which fixes the depth of the NV layer) and the type of surface (polished or as grown) have little impact. However, despite  enhanced spin properties at lower doses, the sensitivity to magnetic fields is essentially independent from the dose because the latter also sets the number of NV centres that contribute to the signal. We then showed that the spin properties can be dramatically improved by a post-implantation annealing at a temperature of 1100-1200$^\circ$C instead of the lower temperature typically used (950$^\circ$C or less). By performing EPR spectroscopy ($T_1$-based and DEER) using the NV layer as a probe, we observed that this higher temperature annealing greatly suppresses the signatures of multivacancy chains visible with the 950$^\circ$C annealing, which implies an overall reduction in the density of residual paramagnetic defects and explains the improvement in the ODMR linewidth and spin coherence time. We also examined the effect of oxygen annealing and observed a modest improvement in the magnetic sensitivity of most NV layers, but a deleterious effect on the samples implanted at the lowest dose. Our EPR spectroscopy measurements revealed the appearance of a broadband magnetic noise (extending over at least 1 GHz) upon oxygen annealing, which disappears after acid cleaning. This suggests that the surface preparation method is a key consideration for these samples depending on the imaging modality ($T_1$ or $T_2$ based).  Finally, the average NV depth was determined by proton NMR measurements, ranging from 10 to 17 nm for implantation energies from 4 to 6 keV, in agreement with previous studies. These results elucidate the parameters that limit the magnetic sensitivity of dense layers of near-surface NV centres, which are used in a growing number of sensing and imaging applications.

\section*{Acknowledgements}

This work was supported in part by the Australian Research Council (ARC) under the Centre of Excellence scheme (project No. CE110001027). L.C.L.H. acknowledges the support of an ARC Laureate Fellowship (project No. FL130100119). J.-P.T acknowledges support from the ARC through the Discovery Early Career Researcher Award scheme (DE170100129) and the University of Melbourne through an Establishment Grant and an Early Career Researcher Grant. D.A.B and S.E.L are supported by an Australian Government Research Training Program Scholarship. T.T acknowledges the support of Grants-in-Aid for Scientific Research from the Ministry of Education, Culture, Sports, Science, and Technology, Japan (No. 15H03980, 26220903, and 16H06326).
 
\bibliographystyle{apsrev4-1}
\bibliography{bib}	
	   
\end{document}